\documentclass{article}
\usepackage[top=1in,bottom=1in,left=1in,right=1in]{geometry}  

\usepackage[utf8]{inputenc}
\usepackage{graphicx, xcolor, amsmath}
\usepackage{setspace}
\usepackage[colorlinks,allcolors=cyan!70!black]{hyperref}

\usepackage{enumitem}

\usepackage{wrapfig,lipsum,booktabs}

\usepackage{multicol,lipsum}
\usepackage{multirow}
\usepackage{hhline}

\usepackage{amsmath}
\usepackage{amsfonts}
\usepackage{amsthm}
\usepackage{amssymb}
\usepackage{mathrsfs}
\usepackage{bbm}
\DeclareMathOperator*{\argmax}{argmax}
\DeclareMathOperator*{\argmin}{argmin}

\renewcommand{\mathbbm}[1]{\text{\usefont{U}{bbm}{m}{n}#1}}

\usepackage{algorithm}
\usepackage{algpseudocode}

\usepackage{caption} 
\newcommand\corrauthor[1]{%
  \begingroup
  \renewcommand\thefootnote{}\footnote{#1}%
  \addtocounter{footnote}{-1}%
  \endgroup
}

\begin{document}

\begin{center}
\begin{spacing}{1.5}
\textbf{\Large AlignOT: An optimal transport based algorithm for fast 3D alignment with applications to cryogenic electron microscopy density maps}
\end{spacing}

\vspace{5mm}
Aryan Tajmir Riahi$^{1}$, Geoffrey Woollard$^{1}$, Frédéric Poitevin$^{2}$, Anne Condon$^{1}$, and Khanh Dao Duc$^{1,3,*}$ \corrauthor{$^*$ corresponding author: kdd@math.ubc.ca}

\vspace{5mm}
$^{1}$Department of Computer Science, University of British Columbia, Vancouver, BC V6T 1Z4, Canada\\
$^{2}$ SLAC National Accelerator Laboratory, Menlo Park, CA 94025, USA\\
$^{3}$Department of Mathematics, University of British Columbia, Vancouver, BC V6T 1Z4, Canada\\

\vspace{5mm}
\vspace{5mm}
\end{center}

\begin{abstract}
Aligning electron density maps from Cryogenic electron microscopy (cryo-EM) is a first key step for studying multiple conformations of a biomolecule. As this step remains costly and challenging, with standard alignment tools being potentially stuck in local minima, we propose here a new procedure, called \emph{AlignOT}, which relies on the use of computational optimal transport (OT) to align EM maps in 3D space. By embedding a fast estimation of OT maps within a stochastic gradient descent algorithm, our method searches for a rotation that minimizes the Wasserstein distance between two maps, represented as point clouds. We quantify the impact of various parameters on the precision and accuracy of the alignment, and show that \emph{AlignOT} can outperform the standard local alignment methods, with an increased range of rotation angles leading to proper alignment. We further benchmark \emph{AlignOT} on various pairs of experimental maps, which account for different types of conformational heterogeneities and geometric properties. As our experiments show good performance, we anticipate that our method can be broadly applied to align 3D EM maps.

\end{abstract}

\section*{Introduction}
Solving the 3D structures of biomolecules is key to their function and the mechanisms underlying biological processes. For this purpose, cryogenic electron microscopy (cryo-EM) 
has become in recent years the most used technique to solve structures \cite{bai2015cryo}. One main advantage of this technique, in contrast with X-ray crystallography, is that it potentially allows various conformations (or 3D configurations) of the same molecule to be solved \cite{poitevin2020structural}. Once different conformations are obtained as 3D EM density maps (i.e., large 3D grids of voxels with different levels of intensities), aligning these maps in 3D space is needed to further compare them. 

Efficient methods have been developed to align two protein structures \cite{mukherjee2009mm,moreta2019probabilistic}, assuming their atomic composition is known. In this case, aligning two conformational structures is tantamount to finding an optimal rigid body transformation (i.e. a combination of 3D translation and rotation) that can align homologous atoms. However, when density maps are only given, one cannot directly establish such a homology correspondence from voxel to voxel, so the same problem becomes more challenging as the grid size increases and with the computational cost of searching over all possible rigid body transformations.

To solve the rigid body alignment problem for 3D cryo-EM density maps, standard approaches use various algorithms to maximize correlation \cite{pettersen2004ucsf, evans2008introduction, kawabata2008multiple}. More recently, Han et al. introduced a new method, which relies on representing the maps as sets of unit vectors before performing alignment  \cite{han2021vesper}. Overall, both the choice of the metric to optimize, as well as the representation of the maps, can play important roles in getting a successful alignment. In this paper, we introduce a novel approach, called \emph{AlignOT}, that uses a point-cloud representation of 3D maps, and minimizes the so-called Wasserstein distance between two maps with a stochastic gradient algorithm. This non-Euclidean distance is associated with the theory of Optimal Transport (OT) \cite{peyre2019computational}, with recent advances that make tractable the computation of transport-based distances \cite{cuturi2013sinkhorn,altschuler2017near}. After describing the procedure of \emph{AlignOT} in detail, we present the results of our experiments that quantify the precision and accuracy of this method, and benchmark it on a set of representative experimental maps. Overall, the good performance of \emph{AlignOT} with respect to standard local alignment methods suggests that our method can be broadly applied, as an alternative for aligning 3D EM maps with a non-Euclidean metric. We finally discuss the potential limitations of \emph{AlignOT}, and connections with other recent methods and problems in optimal transport that would help to further improve and generalize it.

\section*{Material and Methods}
    
    \subsection*{Point cloud representation of EM maps}\label{sec:point_cloud}

To represent voxelized cryo-EM maps, we use the topology representing network algorithm (TRN)  \cite{martinetz1994topology}, which reduces a map of $d^3$ voxels to a point cloud, i.e., a set of $n$ points $\in \mathbb{R}^3$. Briefly, the voxelized maps (typically $\sim100^3$ to $\sim500^3$ voxels in the 3D grid) are first thresholded with a noise floor, to set low intensity regions to zero, and normalized to a probability mass function $\mathcal{P}$ over the grid points. $n$ points $\left({\bf r}_i(0)\right)_{i=1\ldots n}$ are initially sampled from $\mathcal{P}$, and then updated in parallel for $t_f$ rounds by taking weighted steps, according to the following equations \cite{martinetz1994topology} 

\begin{align}
    \label{eq:trn_update}
    {\bf r}_i(t+1) &= {\bf r}_i(t) + \varepsilon(t) \exp[-k_i/\lambda(t)] ({\bf r}_t-{\bf r}_i(t)), \\
    \label{eq:trn_update_lambda}
    \lambda(t) &= \lambda_0 \Big(\frac{\lambda_f}{\lambda_0}\Big)^{t/t_f}, \\
    \label{eq:trn_update_epsilon}
    \varepsilon(t) &= \varepsilon_0 \Big(\frac{\varepsilon_f}{\varepsilon_0}\Big)^{t/t_f},
\end{align}

where ${\bf r}_t$ is a single grid point sampled from $\mathcal{P}$, $k_i$ denotes the rank of point ${\bf r}_i(t)$, by its Euclidean distance to ${\bf r}_t$, and 
$\lambda_0,\varepsilon_0,\lambda_f,\varepsilon_f,t_f$ are hyperparameters. 
In practice, we used
 $\varepsilon_0=0.3$ and $\varepsilon_f=0.05$ for the initial and final step sizes, $\lambda_0=0.005 \times n$ and $\lambda_f=0.5$ for the initial and final scaling ranks, and $t_f=8 \times n$ for the total number of steps (adapted from \cite{trn2021}).

     \subsection*{Optimal Transport and Wasserstein distance}\label{sec:OT}
To compare the point cloud representations of EM maps, we use a non-Euclidean metric that derives from the theory of Optimal Transport \cite{peyre2019computational}. For two given point clouds, $\mathbf{A} = \{a_1,\dots,a_n\}$ and $\mathbf{B} = \{b_1, \dots, b_n\}$, we define a cost matrix $C_{i,j} =
d(a_i,b_j)^2$, where $d$ is the Euclidean distance. 
The \emph{entropy regularized $2$-Wasserstein distance} between $\mathbf{A}$ and $\mathbf{B}$, denoted by $\mathcal{W}_{2,\epsilon}(\mathbf{A},\mathbf{B})$, is then defined as 

\begin{equation}
\begin{aligned}
\mathcal{W}_{2,\epsilon}(\mathbf{A},\mathbf{B}) = [\min_{P \in \mathbb{R}_+^{n \times n}} \quad & \ \sum_{i,j=1}^n C_{i, j}P_{i, j} + \epsilon H(P)]^{1/2}\\
\textrm{s.t.} \quad & P.\mathbbm{1} = P^T.\mathbbm{1} = \mathbbm{1}/n\\
\end{aligned}
,
\label{eq:Wass_dist}
\end{equation}
where $\epsilon \in \mathbb{R}_+$ is the \emph{regularization parameter} and the entropy $H(P)$ is given by
\begin{equation}
    H(P) = \sum_{i,j=1}^n P_{i,j}\log P_{i,j}.
\end{equation}

The minimizer of equation \eqref{eq:Wass_dist} is called the \emph{transport plan}. For the rest of the Methods section, we will simply denote the Wasserstein distance as $\mathcal{W}_{2,\epsilon}$, and $P_{i(a),i(b)}$ as $P_{a,b}$, where $i(a)$ and $i(b)$ are the indices of the two points $a$ and $b$ in $\mathbf{A}$ and $\mathbf{B}$, respectively.

\subsection*{\emph{AlignOT}: Algorithm for 3D map alignment}\label{sec:alg}
To align two 3D maps from their point cloud representations $\mathbf{A}$ and $\mathbf{B}$, we solve the optimization problem  
        \begin{equation}
            \label{eq:minOT}
            q_\text{opt} =
            \argmin_{q \in \mathbb{H}} \mathcal{W}_{2,\epsilon}(R_q(\mathbf{A}), \mathbf{B}),
        \end{equation}
where $q$ is a quaternion (defined over the quaternion space $\mathbb{H}$), that we identify to a 3D rotation $R_q$ in $SO(3)$, so $R_q(\mathbf{A}) = \{R_q(a_i) | a_i \in \mathbf{A}\}$. We explain in the Results section 
and Appendix A  
why we can only consider rotations and ignore translations 
to solve the general alignment problem, and provide more details on the identification of $q$ to $R_q$ in Appendix B. 
Our stochastic gradient descent procedure to solve \eqref{eq:minOT}, called \emph{AlignOT}, is detailed in Algorithm \ref{alg2}. At each iteration, the algorithm updates $q$ from the transport plan $P$ between $R_q(\mathbf{A})$ and $\mathbf{B}$ as follows: After sampling one point $a \in R_q(\mathbf{A})$, we evaluate $\pi(a) = \argmax_{b \in \mathbf{B}} P_{a,b}$, and compute the gradient in $q$ associated with $d(\pi(a), a)^2$, where $d$ is the Euclidean distance, to update $q$. To compute the transport plan, we apply the Sinkhorn algorithm \cite{cuturi2013sinkhorn}, with the initial vectors set as the outputs of the previous iteration. In practice, we also set the convergence condition $\|d(\pi(a) , a)^2\| < \delta$ (where $\delta>0$), that stops the algorithm before the maximum number of iterations. The hyperparameters of this procedure are the learning rate $\alpha$ associated with gradient descent, the regularization parameter $\epsilon$ associated with the Wasserstein distance, and a threshold $\delta$ associated with the number of iterations. In all our experiments, we set $\epsilon = 100$, $\delta = 10^{-10}$, and the maximum number of iterations equal to $500$. 
    
	\begin{algorithm}
		\caption{\emph{AlignOT}: 3D density maps alignment with SGD using unit quaternions and Wasserstein distance}
		\label{alg2}
    		\hspace*{\algorithmicindent} \textbf{Input} 
    		two 3D density maps $\mathcal{A}, \mathcal{B}$, number of sampled points $n \in \mathbb{R}$, learning rate $\alpha \in \mathbb{R}$, regularization parameter $\epsilon \in \mathbb{R}$, maximum number of iterations $L \in \mathbb{N}$, and gradient threshold $\delta \in \mathbb{R}$
		\begin{algorithmic}[1]
            \State Sample two sets of $n$ points $\mathbf{A},\mathbf{B} \subset \mathbb{R}^3$ from $\mathcal{A}, \mathcal{B}$ respectively, using TRNs
			\State $q = 1 + 0i + 0j + 0k$
			\State $G = \alpha^2$
            \While {not converged \textbf{and} the number     of iterations is at most $L$} 
            \State Compute $R_q(\mathbf{A})$
            \State Compute $P$ to be the OT plan matrix between $R_q(\mathbf{A})$ and $\mathbf{B}$
            \State Randomly select $a\in R_q(\mathbf{A})$
            \State $b = \pi(a)$
            \State $f(q) = 
            d(R_q(a) , b)^2$ (where $d$ is Euclidean distance in $\mathbb{R}^3$)
            \State $G = G + \|\nabla f(q)\|^2$
            \State $q = q - \frac{\alpha}{\sqrt{G}} \times \nabla f(q)$
            \State $q = \frac{q}{\|q\|}$
			\EndWhile
			\State \Return $q$
		\end{algorithmic}
	\end{algorithm}

\subsection*{Implementation}\label{sec:implementation}
We implemented \emph{AlignOT} in Python 3.6.4. To sample a point cloud representation of an EM map using TRN, we adapted code from ProDy \cite{prody2021}. We used the NumPy package for matrix operations and POT's implementation of the Sinkhorn algorithm, which was modified to set the initial vectors (instead of initializing with uniform vectors). Our code is available in this \href{https://github.com/artajmir3/RPE}{GitHub repository}.

\subsection*{Datasets}\label{sec:dataset}

\begin{figure}[h!]
\centering
    \includegraphics[width=.99\textwidth]{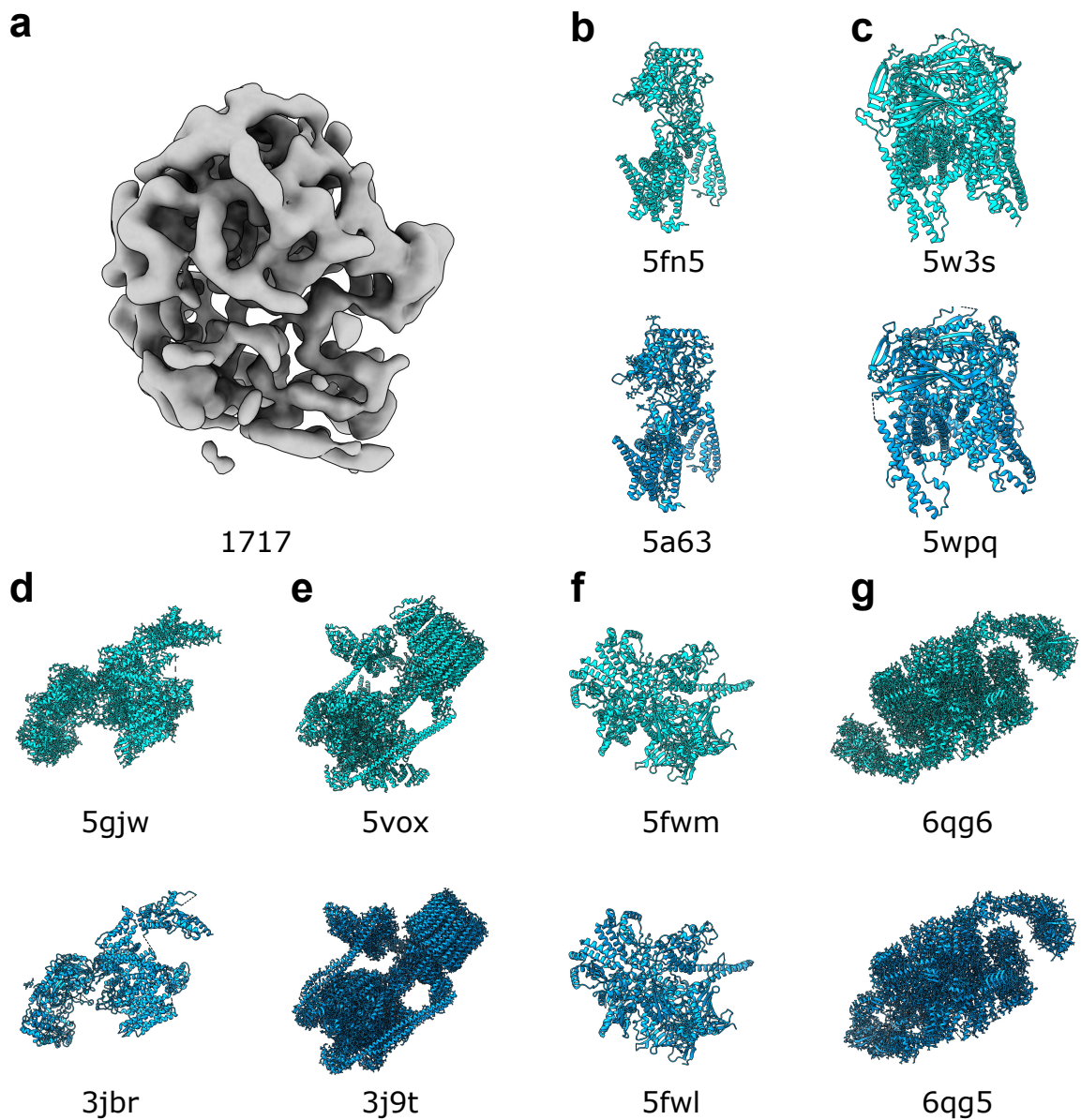}
  \caption{3D maps used in our experiments, visualized with Chimera \cite{pettersen2004ucsf}. \textbf{(a):} Density map used in our first experiment (EMDB:1717). \textbf{(b-g):} Pairs of maps used in the second set of experiments, representing different conformational states of a given complex/molecule (see also Table \ref{tab:dataset2}). }
  \label{fig:maps}
\end{figure}

We tested \emph{AlignOT} on various publicly available EM maps available from the EMDB, as listed in Table \ref{tab:dataset2}, and shown in Figure \ref{fig:maps}. For experiments that involved aligning a pair of distinct maps, we also used the corresponding pair of structures taken from the PDB \cite{berman2007worldwide}, and used \emph{MM-align} \cite{mukherjee2009mm} to set a ground truth alignment, before converting the structures into EM maps using the function \texttt{molmap} in Chimera X \cite{goddard2018ucsf}. All the datasets used in this study are available at this \href{https://osf.io/typ9s/}{OSF page}.

\section*{Results}

\subsection*{Formulation of the alignment problem for EM maps with point cloud representation}\label{sec:formulation}
We present here a new procedure, called \emph{AlignOT}, that aligns 3D density maps from cryo-EM. Assuming that the EM maps are represented by two 3D point clouds $\mathbf{A} = \{a_1,\dots,a_n\}$ and $\mathbf{B} = \{b_1,\dots,b_n\}$, and for a given distance function $d$ defined over the space of point clouds, the problem of aligning these maps consists of  finding a rigid body transformation that minimizes the objective function
 
        \begin{equation} \label{eq:loss_def}
         \mathcal{L}_d(R, T) = d(\text{move}_{R,T}(\mathbf{A}),\mathbf{B})^2,
        \end{equation}
  where $\mathcal{L}_d(R, T)$ is defined over rotation matrices $R \in SO(3)$ and translation vectors $T \in \mathbb{R}^3$, and the operator $\text{move}_{R,T}(\mathbf{A})$ is defined as 
    \begin{equation}\label{eq:move}
        \text{move}_{R,T}(\mathbf{A}) = \{Ra_i + T| a_i \in \mathbf{A}\}.
        \end{equation}
        
As the choice of $d$ influences both the accuracy and the computational cost of the solution to the rigid body alignment problem, we here use the 2-Wasserstein distance, associated with the theory of Optimal Transport \cite{peyre2019computational}. This distance can be used to compute distances between probability distributions, and is applied here more specifically for two distributions of 3D point clouds of same size. To efficiently evaluate this distance, we consider a regularized version (see equation \eqref{eq:Wass_dist} in the Methods section 
), denoted $\mathcal{W}_{2,\epsilon}$. Besides, given the centers of mass $\Bar{a} = \frac{1}{n}\sum_{i=1}^n a_i$ and $\Bar{b} = \frac{1}{n}\sum_{i=1}^n b_i$, and the centered point clouds $\mathbf{A_c} = \{a_{c_i} = a_i - \Bar{a}|a_i \in \mathbf{A}\}$ and $\mathbf{B_c} = \{b_{c_i} = b_i - \Bar{b}|b_i \in \mathbf{B}\}$, we can show that the optimal translation of the objective function \eqref{eq:loss_def} is

\begin{equation}
    \label{eq:opt_translation}
        T_\text{opt} = \Bar{b} - R_{\text{opt}}\Bar{a},
    \end{equation} where
        
\begin{equation}
    \label{eq:minOT_new}
    R_\text{opt} =
    \argmin_{R \in SO(3)} \mathcal{W}_{2,\epsilon}(R(\mathbf{A_c}), \mathbf{B_c}).
\end{equation}
      
Thus, the search for an optimal rigid body transformation in \eqref{eq:minOT_new} can be simplified to rotations after matching the centers of mass of $\mathbf{A}, \mathbf{B}$, leading to equation \eqref{eq:minOT} of the Methods section 
. We provide a detailed proof of this result in Appendix A. 

\subsection*{\emph{AlignOT} optimization procedure and complexity}

To search for an optimal rotation that minimizes the Wasserstein distance between two 3D point clouds, we use an iterative Stochastic Gradient Descent algorithm, detailed in Algorithm \ref{alg2}. Basically, the algorithm aims to improve the alignment at each iteration, by rotating a map according to an assignment provided by the evaluation of the Wasserstein distance, with quaternions used to represent the 3D rotations (see Appendix  B 
for a full description). Since EM maps are defined on a 3D voxelized grid, we also first convert the maps into point clouds, using the TRN algorithm \cite{martinetz1994topology}. We cover the application of the TRN application, the Wasserstein distance and the optimization procedure, as well as its implementation in the Material and Methods sections 
.

The complexity of our method (summarized in Algorithm \ref{alg2}) can be evaluated as follows: Assume that $L$ is the maximum number of iterations, $n$ the size of the point cloud, and $\epsilon$ the regularization parameter of the $\mathcal{W}_{2,\epsilon}$ distance. Each iteration of this algorithm consists of three steps. First, rotating one point cloud, second, computing the OT plan matrix, and last, sampling a random point and computing the gradient. Rotating a point cloud requires computing the coordinates of each point after rotation and takes $O(n)$ time, where $n$ represents the number of points. To compute the OT plan matrix we use the Sinkhorn algorithm \cite{altschuler2017near} which solves this problem in $O(n^2 \log n \epsilon^{-3})$ time, where $\epsilon$ is the regularization parameter. Finally, the gradient step takes $O(1)$ time. Overall, the most time-consuming part of this algorithm is computing the OT plan matrix at each iteration. Therefore, the overall time complexity of the algorithm is $O(n^2L \log n \epsilon^{-3})$.
\subsection*{Alignment between maps can be obtained by minimizing the Wasserstein distance}\label{sec:exp1}

\begin{figure}[h!]
\includegraphics[width=0.99\textwidth]{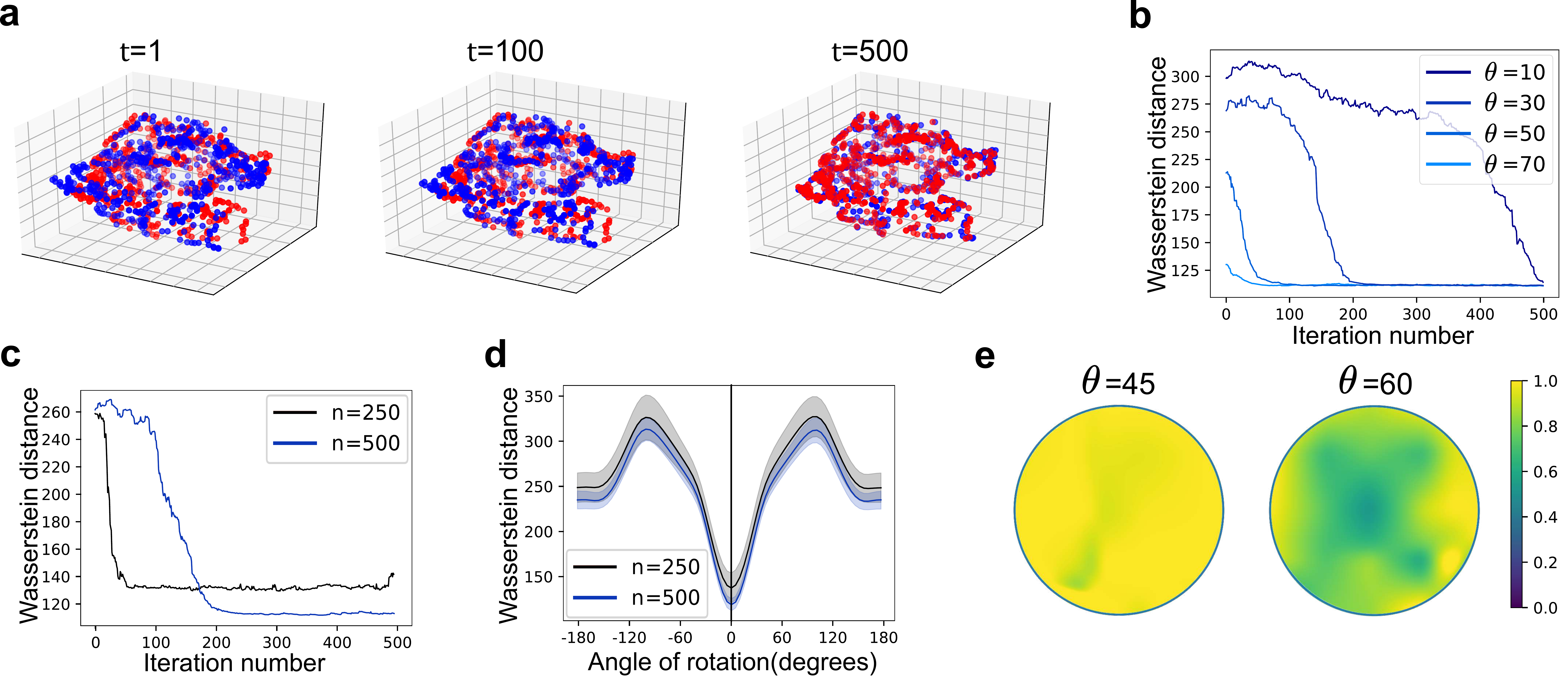}
\centering
\caption{Alignment of two copies of a single map (from EMDB 1717) using \emph{AlignOT}  \textbf{(a):} We generated two copies of the map that differ from a 3D rotation, and applied \emph{AlignOT} (see Algorithm \ref{alg2}). The figure shows the result of the procedure at different iterations  ($t=1, 100, 500$). The blue and red dots represent the rotated and target point clouds, respectively.  \textbf{(b):} For different initial values $\theta$ ($= 10^\circ, 30^\circ, 50^\circ, 70^\circ$) of the angle difference between the two maps and same rotation axis, we plot the Wasserstein distance between the two target and rotated point clouds across the iterations of the algorithm, showing that they all converge to the same limit. \textbf{(c):} For the same initial rotation (with $\theta= 50^\circ$), we plot the variations of the Wasserstein distance across iterations as in \textbf{(b)}, for point clouds of size $n=250$ (black) and $500$ (blue). 
\textbf{(d):} For the same rotation axis as in \textbf{(b)} and \textbf{(c)}, we plot the average Wasserstein distance between two sampled point clouds as a function of $\theta$, for point clouds of size $250$ (black) and $500$ (blue). Error regions show the standard deviation from sampling 100 different point clouds for each angle. \textbf{(e):} Alignment success rate of \emph{AlignOT} at fixed angles $\theta = 45$, and $60^\circ$, and over the rotation axes that cover the upper hemishpere of $S^2$. Heatmaps show the percentage of outcomes that result in an alignment with error $\leq 5^\circ$, where each point of the disk is the projection of the axis considered in $S^2$ (89 in total, with 20 runs for each).
}
\centering
    \label{fig:comp1}
\end{figure}

To evaluate our method, we first tested \emph{AlignOT} on aligning two point clouds that differ by a rotation only. To do so, we used  an experimental map of a ribosome from EMDB 1717,  shown in Figure \ref{fig:maps}a (ribosome structures are broadly studied in cryo-EM \cite{poitevin2020structural,kushner2022riboxyz}). First, we sampled two clouds of 500 points, and applied a rotation defined in its axis-angle representation by an arbitrary axis, and an angle $\theta = 20^\circ$. Figure \ref{fig:comp1}a illustrates how the moving point cloud gets closer to the targeted one over the iterations of the algorithm, until the convergence criterion is reached. 
To confirm this visual impression, we repeated the procedure with different initial angles $\theta \in \{10^\circ, 30^\circ, 50^\circ, 70^\circ\}$. The corresponding Wasserstein distance obtained across the iterations is shown in Figure \ref{fig:comp1}b, with all the four trajectories converging to the same value and resulting in a successful alignment. However, we also observed that as $\theta$ increases, it takes more iterations for the algorithm to converge, with longer periods of slow variations at the beginning of the procedure, suggesting that this alignment can only be achieved within a certain range of $\theta$. Aside from the initial angle, we also studied in Figure \ref{fig:comp1}c how the size of the point cloud can affect the convergence plot. For $\theta= 50^\circ$, decreasing the point cloud size from 500 to 250 leads to a potential loss of precision (with the Wasserstein distance between two aligned point clouds increasing from $\sim$ 
115 to 130), but with faster convergence from the algorithm (from $\sim$ 200 to 50 iterations to converge), indicating some trade-off between accuracy and speed. 

\begin{figure}[h!]
     \centering
      \includegraphics[width=.8\textwidth]{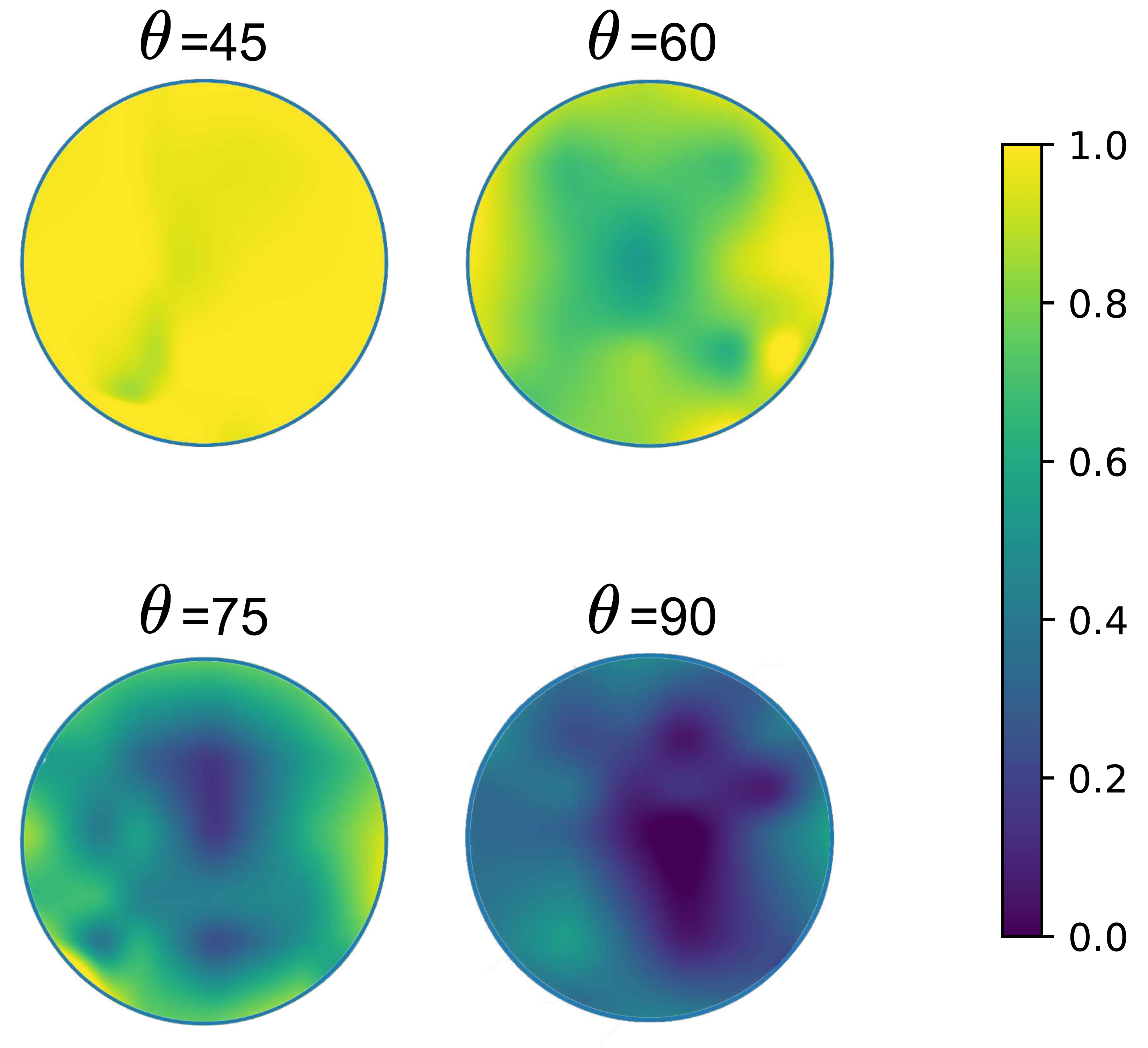}
     \caption{Alignment success rate of \textit{AlignOT} at fixed angles $\theta = 45, 60, 75$, and $90^\circ$, and over the rotation axes that cover the upper hemishpere of $S^2$. Heatmaps show the percentage of outcomes that result in an alignment with error $\leq 5^\circ$, where each point of the disk is the projection of the axis considered in $S^2$ (89 in total, with 20 runs for each).}
    \label{fig:s1}
 \end{figure}

To interpret these results, 
we further plotted in Figure \ref{fig:comp1}d how the Wasserstein distance varies on average (after sampling different point clouds of same size 250 and 500), as a function of $\theta$ (and same rotation axis). 
In addition to the global minimum achieved for $\theta=0$, another local minimum was detected at $\theta=180^\circ$, separated by a peak around $90^\circ$. While the sharp decrease observed towards the global minimum suggests that the optimization procedure can converge well for $\theta \leq 60^\circ$, it is also possible that the algorithm does not converge to the global minimum above this range. Besides, the higher variability obtained from sampling point clouds of size 250, compared with 500, suggests that the final alignment may be less accurate in this case. While the same fixed rotation axis was considered in the previous experiments (with different values of $\theta$), we finally evaluated the probability to successfully align the maps for initial rotations of fixed angle $\theta$ (45, 60, 75 and 90$^\circ$), and across different axes covering half of the sphere $\mathcal{S}^2$. Upon mapping the axes on the planar disk in Figure \ref{fig:comp1}e, we found local regions of poorer alignment that match with local minima of the Wasserstein distance. These results also confirm the existence of a limiting range within which the method can align two maps. While a successful alignment is overall obtained for $\theta=45^\circ$, the maps get partially aligned in different regions of the disks for $60^\circ$ (see  Figure \ref{fig:comp1}e), with the performance worsening as $\theta$ increases (see Figure \ref{fig:s1}).

 \subsection*{Benchmarking performance and accuracy of \emph{AlignOT} with a pair of identical maps}  \label{sec:benchmark}
 
 We next focused on point cloud size and the range of convergence, and quantified their impact on the accuracy and computational cost of \emph{AlignOT}. We first considered the same map and alignment task with $\theta = 20^\circ$ as previously, with  different point cloud sizes $n$ (from 50 to 1000). The results obtained with a standard workstation are shown in Table \ref{tab:numpoints}, and confirm the existence of a trade-off between accuracy and speed, with the alignment improving as $n$ increases, but with a larger runtime that goes from a few seconds for $n=50, 100, 200$ to approximately a minute for $n=1000$. While the accuracy of the alignment remains  poor for $n=50$ and 100 (with an average error of $12.6^\circ$ and $7.72^\circ$ respectively), it significantly improves  for $n=500$ with $2.35 \pm 1.23^\circ$ error observed, with a runtime suggesting that \emph{AlignOT} can  be used in practice on standard density maps. We also noted that \emph{AlignOT} runs slower in comparison with Chimera and Chimera X's alignment function \texttt{fitmap}, which performs a 
 steepest ascent optimization
 to align maps according to their overlapping score \cite{pettersen2004ucsf,goddard2018ucsf} ($0.3$ s on average). On the other hand, we show in our next experiments  that \emph{AlignOT} outperforms \texttt{fitmap} in accuracy as $\theta$ increases. 

\begin{figure}[h!]
     \centering
      \includegraphics[width=.8\textwidth]{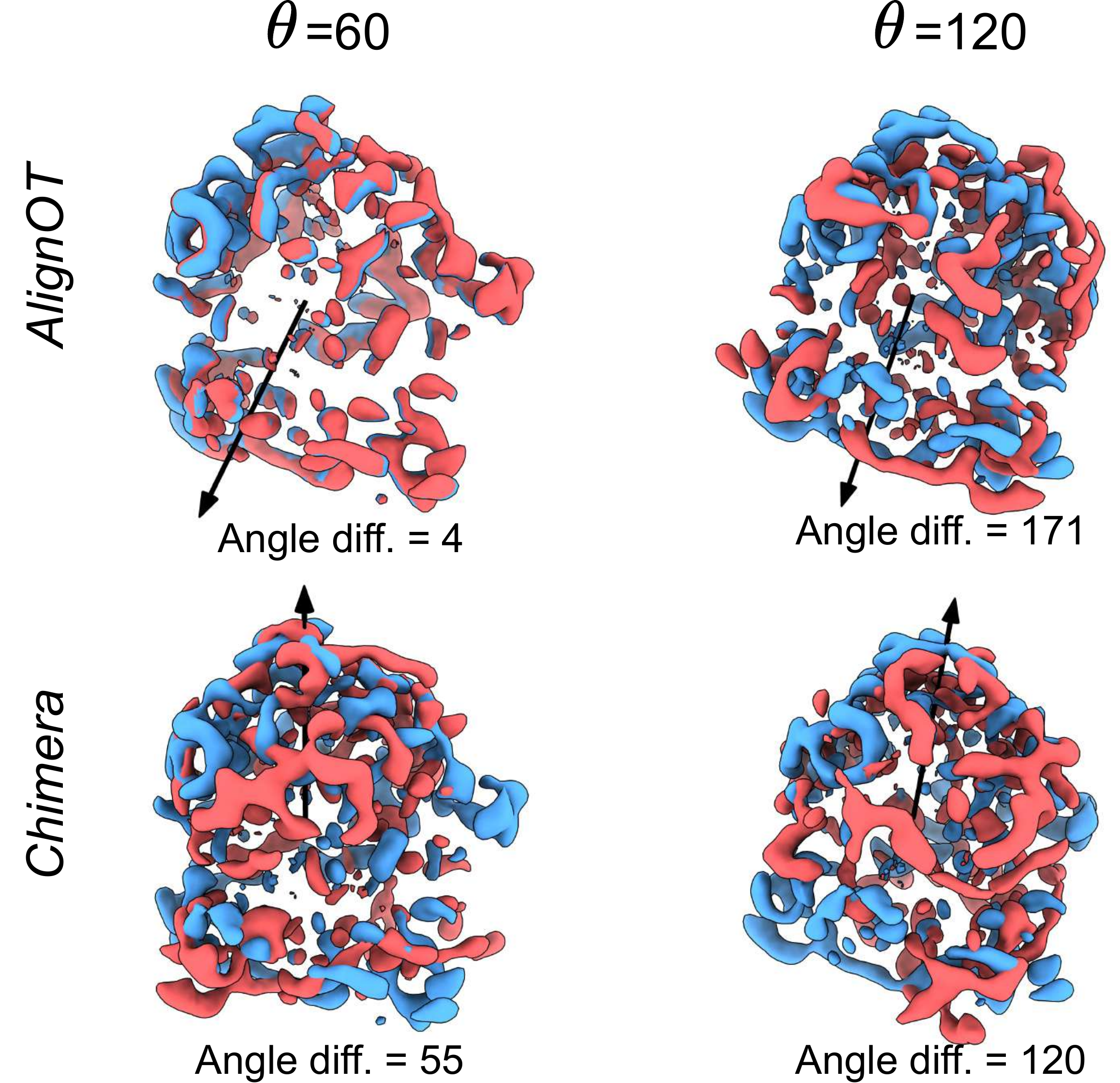}
     \caption{Visual comparison between \textit{AlignOT} and Chimera's \texttt{fitmap} function. We set up a couple of maps from EMDB:1717  that differ from a rotation of fixed axis and angles $\theta=60$ and $120^\circ$,  and ran \textit{AlignOT} and Chimera's \texttt{fitmap} function local search. The figure shows the result of the procedure, with the final rotation (axis, angle) between the two maps.}
 \label{fig:s2}
 \end{figure}

 More precisely, we 
  determined the range of $\theta$ within which the method converges, and compared our method with 
 \texttt{fitmap}. 
 As shown in Figure \ref{fig:comp1}c, we found that \emph{AlignOT} can cover a wider range that extends to $75^\circ$, while \texttt{fitmap} starts failing at approximately $45^\circ$  (see also Figure \ref{fig:s2} for a visualization of some representative alignments obtained). Beyond this value, \emph{AlignOT} leads to some variable results up until $100^\circ$, due to the stochasticity of the algorithm. It then converges towards the other local minimum found for the Wasserstein distance observed in Figure \ref{fig:comp1}d, with a difference of $\sim 175^{\circ}$ from the true alignment. 
  In practice, Chimera also provides a global search option that randomly generates different initial instances of the rotated map, and keeps the best alignment to the target map using \texttt{fitmap}. To estimate the potential gain from using \emph{AlignOT}, we generated $500$ random initial placements, and computed the rates of successful alignments from \emph{AlignOT} and \texttt{fitmap}'s local search, to be $15.6\%$ and $2.2\%$, respectively. 
 In this context, \emph{AlignOT} thus reduces the number of initial placements needed in the global search by a factor of $7.6$ (as $(1-0.022) ^ {7.6} \approx 1-0.156$).

    \subsection*{Benchmarking \emph{AlignOT} with conformationally heterogeneous pairs}
    
\begin{figure}
\centering
    \includegraphics[width=0.99\textwidth]{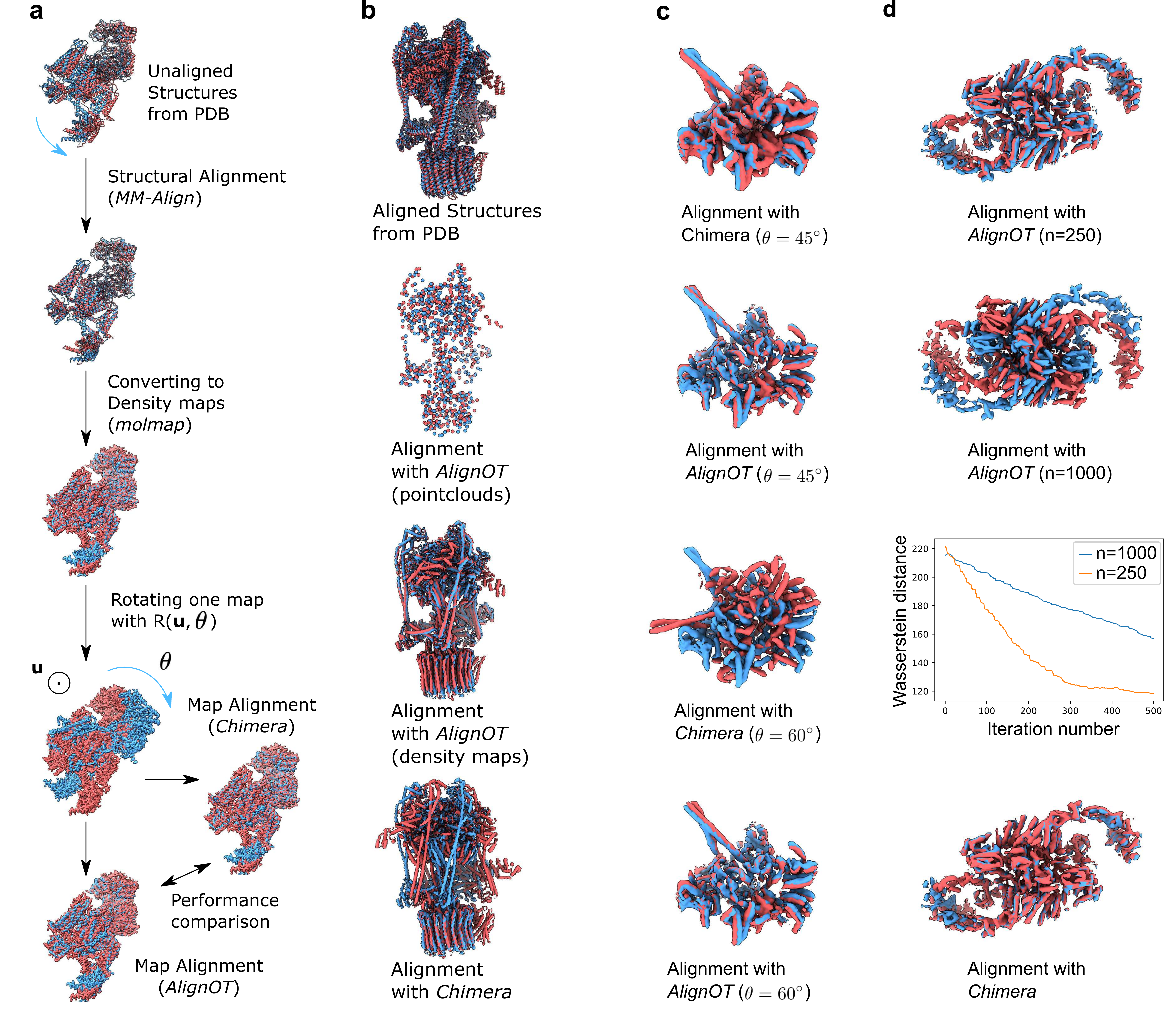}
  \caption{
  Performance of \emph{AlignOT} with conformationally heterogeneous pairs \textbf{(a):} To quantify the performance of \emph{AlignOT} on aligning different maps, we first aligned the corresponding PDB structures using \emph{MM-align} \cite{mukherjee2009mm} to set a ground truth alignment, and converted the aligned structures into cryo-EM density maps using the \texttt{molmap} command in Chimera \cite{goddard2018ucsf}, which were then used to benchmark 
  Chimera's \texttt{fitmap} function and \emph{AlignOT} with different initial rotations (see also Table \ref{tab:exp_benchmark}). \textbf{(b):} Alignments obtained for
  the pair ID 4 (see Table \ref{tab:dataset2})  and  initial rotation angle $\theta=45^\circ$. From top to bottom, we compare the alignment obtained with (1) \emph{MM-Align} from PDB structures, (2) \emph{AlignOT} using a point cloud size $n=500$, with both point clouds and  resulting density maps shown, and (3) Chimera's \texttt{fitmap} function (from the density maps).  \textbf{(c):} Alignments obtained with the pair ID 5 show the increased range of convergence of \emph{AlignOT} compared with \texttt{fitmap}. While \emph{AlignOT} successfully align the maps for $\theta=45$ and $60^\circ$ (with the figure showing the result for $n=1000$), it is not the case for \texttt{fitmap}, with a misalignment observed at $60^\circ$. \textbf{(d):} From top to bottom, aligning pair ID 6 with $\theta=45^\circ$ shows a decrease in accuracy for \emph{AlignOT} using a point cloud size $n=1000$, compared with $n=250$. We plot the corresponding the Wasserstein distance over the iterations of the algorithm, showing a larger final value for $n=1000$, as the algorithm converges more slowly. As a result, this is one of the few cases where we found that Chimera produces a better alignment.}
  \label{fig:exp2}
\end{figure}

 We finally tested \emph{AlignOT} on pairs of distinct maps, as this reflects how the method should be applied in practice. More precisely, we considered six pairs, listed in Table \ref{tab:dataset2} and shown in Figure \ref{fig:maps}, that all account for different conformations of a protein or complex structure, and that were recently used to evaluate another alignment method \cite{han2021vesper}. As illustrated in Figure \ref{fig:exp2}a, we first used the molecular structures associated with the maps to define a ground truth alignment using the structure-based alignment function \emph{MM-align} \cite{mukherjee2009mm} (see also the Datasets section 
 ). This ground truth was used to evaluate the performance of \emph{AlignOT}, as well as Chimera's \texttt{fitmap}, which we used for benchmarking against our method.

Upon testing for different values of point cloud size $n \in \{250,500,1000\}$ and initial rotation angle $\theta\in \left\{45^{\circ},60 ^{\circ},90^{\circ} \right\}$, we recorded the angle differences between the ground truth alignment and the output using both \texttt{fitmap} and \emph{AlignOT}, as reported in Table \ref{tab:exp_benchmark}. Over the 18 cases tested, \emph{AlignOT} outperformed \texttt{fitmap} 13 times (72\%), with some significant improvement observed in the majority of them (8 cases with an average improvement of more than 20$^\circ$). For example, the improvement obtained in the pair ID 4 ($\theta=45^\circ$) is visualized in Figure \ref{fig:exp2}b, and shows how the presence of some symmetries (observed in the figure at the bottom of the molecule) can lead \texttt{fitmap} to misalign maps. Our experiments also confirm the improvement of the range of convergence using \emph{AlignOT}, which we illustrate in Figure \ref{fig:exp2}c with the pair ID 5 (with good alignment obtained at $60^\circ$, contrary to \texttt{fitmap}). Besides, over the 5 cases where \texttt{fitmap} outperformed \emph{AlignOT},  either the difference between the two methods is marginal (less than 2$^\circ$, for ID 1, 3, 6, and $\theta= 45^\circ$), or both methods perform poorly (ID 2 with $\theta= 60,90^\circ$). With the exception of the pair ID 2 (with poor performance from both methods), we found that the best alignment yields a Wasserstein distance that is the closest to that of the ground truth alignment, regardless of the method associated with it. These results thus suggest that the Wasserstein distance, which determines the objective function of our method, is a more appropriate metric to use than the Euclidean norm. Finally, our experiments highlight that increasing point size cloud does not necessarily lead to the best alignment from \emph{AlignOT}, as a result of the trade-off between performance and speed that we also previously observed. Such a phenomenon was observed for the pair ID 6 ($\theta=45^\circ$), with a better alignment for $n=250$ than $n=1000$, as illustrated in in Figure \ref{fig:exp2}d. As shown by the Wasserstein distance plots, the convergence is in this case significantly slower for $n=1000$, which leads the algorithm to stop before the convergence condition gets achieved.
\section*{Discussion}
In this paper, we present \emph{AlignOT}, a new method for aligning cryo-EM density maps that relies on minimizing the Wasserstein distance between sampled point clouds. As shown in our experiments,  \emph{AlignOT} is scalable to the typical 
size of density maps, with a good compromise between accuracy and speed that can be achieved upon tuning the point cloud size. 
Our method can thus be used to quickly align maps that come from different conformations of the same protein or complex. In particular, optimizing for a transport-based metric, instead of other common metrics (e.g. overlap, correlation), allows \emph{AlignOT} to generally outperform the standard local optimization method implemented in Chimera. Interestingly, the Wasserstein (or Earth-mover) distance was used in other applications in cryo-EM and tomographic projections (e.g. in interpolation or clustering \cite{ecoffet2020morphot, ecoffet2022application,  zelesko2020earthmover, rao2020wasserstein}), as its natural interpretation as the cost of displacing mass between two distributions makes it appropriate to compare volume-objects. While the choice of the TRN algorithm to generate point clouds is justified by its previous use to represent molecular structures \cite{zhang2021state}, it can also be replaced by any other point cloud generation method
, and it would be interesting to explore how to possibly improve our method on this aspect (in particular, we could for example use Vector Quantization, as it also has been used for approximating EM maps \cite{de2002modeling, jonic2016versatility}). 

In the more general context of solving a rigid body alignment problem, \emph{AlignOT} can be seen as a variant of the Iterative Closest Point method (ICP) framework \cite{besl1992method}, that consists of iteratively moving the point clouds according to the best way to match them. Among the many different variants of the ICP, Grave \emph{et al}. similarly employed the Wasserstein distance to align language models \cite{grave2019unsupervised}. Compared with their approach, our method differs with the use of quaternions in the gradient evaluation, as they make it possible to efficiently represent and manipulate 3D rotations. In addition, our modified version of the Sinkhorn algorithm, which evaluates the transport plan between the point clouds, provides a more rapid computation of the transport plan at each iteration, by simply reusing the vectors estimated at the previous iteration.

While Chimera's \texttt{fitmap} or \emph{VESPER} \cite{han2021vesper} support a global search option (with multiple initializations of the local search function), we have not at this stage implemented a similar option for \emph{AlignOT}, which also limits the direct comparison with these global methods. In addition, it also be interesting to study how to further optimize 
the learning rate and cloud size 
to improve the method. Yet, as our comparative experiments indicate an improvement of the range of convergence, as well as the number of random placements leading to a successful alignment, they suggest that \emph{AlignOT} can potentially provide a useful alternative from the aforementioned current tools, that could also in the future be implemented as a plug-in for Chimera (as in \cite{ecoffet2020morphot}). Another current limitation is that the optimization problem assumes that the maps to align represent different conformations of a molecule, and thus carry approximately the same total density. However, it is also important in the context of Cryo-EM to consider the case (which \emph{VESPER} does), of fitting two maps of different sizes, with one representing only a part of the other \cite{evans2008introduction, kawabata2018rigid}. In this case, since we cannot simplify, as in Appendix  A 
, the rigid body alignment problem into an optimization over the rotations only, our method does  not apply directly. However, as  one can naturally formulate this problem in our framework as a problem of \emph{unbalanced}, or \emph{partial} Optimal Transport \cite{chizat2015unbalanced}, and with the recent development of computational methods to solve it \cite{koehl2021physics,le2022multimarginal}, it makes such a generalization of \emph{AlignOT} a promising future direction to pursue.

\section*{Acknowledgments}
This research was supported by a  NFRFE-2019-00486 grant, and through computational resources and services provided by Advanced Research Computing at the University of British Columbia.

\bibliographystyle{plos2015}
\bibliography{bibli}
\newpage

\section*{Figures}

\newpage

\newpage

 \newpage

 \begin{figure}
 \includegraphics[width=0.7\textwidth]{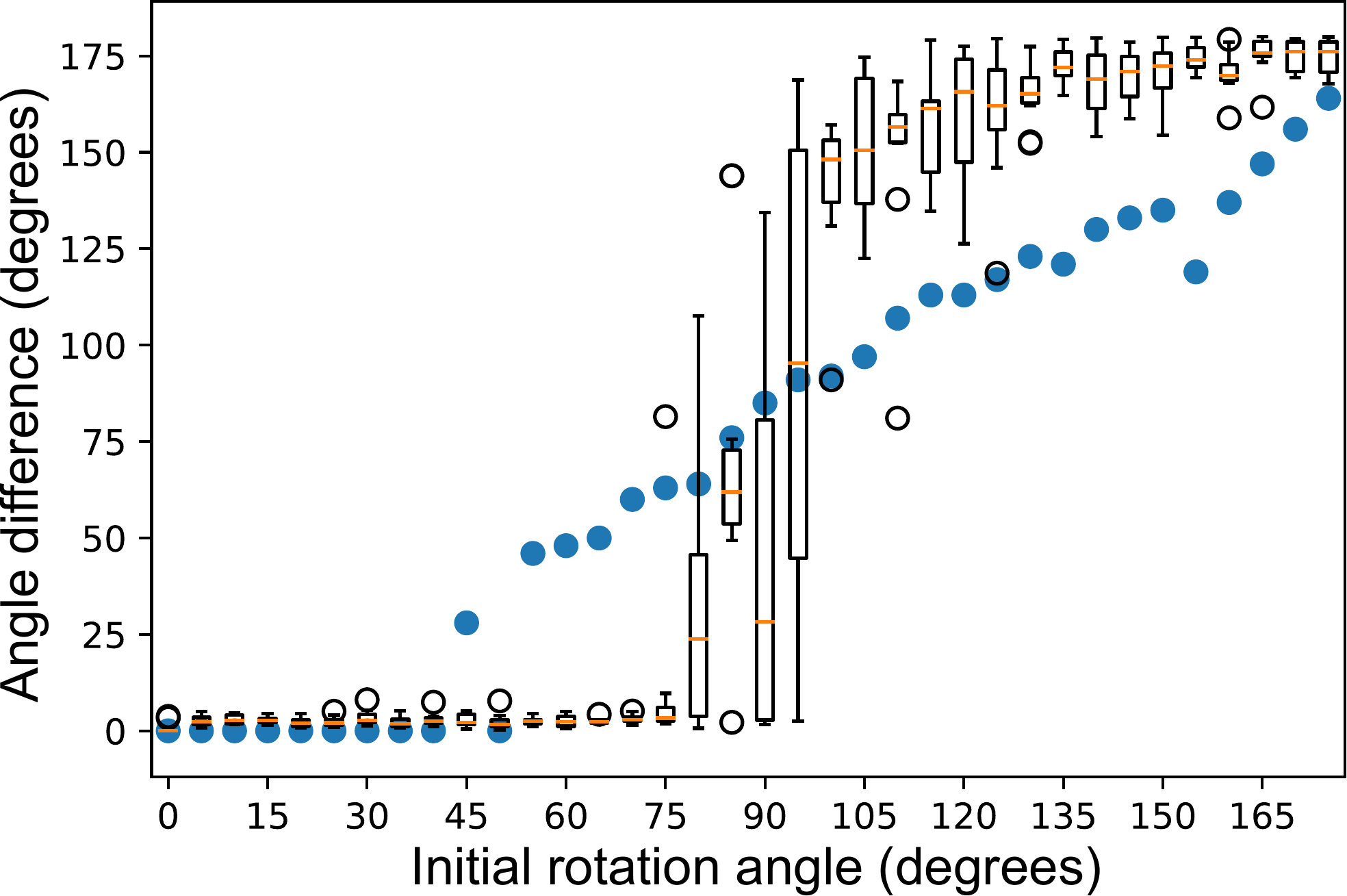}
 \centering
     \caption{Comparison between \emph{AlignOT} and Chimera's \texttt{fitmap} function:  With a fixed axis and for a rotation angle difference $\theta$ between $0$ and $180$ degrees (x-axis), we set up a couple of maps from EMDB:1717 and run \emph{AlignOT} and Chimera's \texttt{fitmap} function local search. The blue dots represent the error obtain in the alignment using \texttt{fitmap}, while the box plot
     (showing the minimum, first quartile, median, third quartile, and maximum)
     indicates the error of \emph{AlignOT} over $10$ runs.}
     \label{fig:comp_chimera}
 \end{figure}
 \newpage

\newpage
 
 \newpage
 
 \clearpage
 
\section*{Tables}

\begin{table}[h!]
\begin{center}
\caption{EM maps used in our experiments, also shown in Figure \ref{fig:maps}. The map from EMDB:1717 is used for aligning two copies of the same map. The other maps are grouped and aligned in our experiments as pair of different conformations of the same molecule, with the PDB structures used to obtain a ground truth (see Datasets and Results sections).}
\label{tab:dataset2}
\resizebox{0.48\textwidth}{!}{
\begin{tabular}{ | c | c | c | }
\hline
 EMDB ID
 & Protein or Complex Name & 
 Figure \\ 
 \hline
  1717 \cite{fischer2010ribosome} &
  Ribosome  &  1 a
                    \\
\hhline{|=|=|=|}
 EMDB/PDB ID
 & Protein or Complex Name & 
 Figure/ID \\ 
 \hline
 3240 / 5fn5 \cite{bai2015sampling} &
 \multirow{2}{*}{Human $\gamma$-secretase} &  
 \multirow{2}{*}{ \ref{fig:maps} b / 1}
                    \\
                     \cline{1-1}
                     2677 / 5a63 \cite{bai2015atomic} & &\\
 \hline
 8881 / 5wpq \cite{chen2017structure} &
 \multirow{2}{*}{TRPML} &
 \multirow{2}{*}{ \ref{fig:maps} c / 2 }
                    \\
                     \cline{1-1}
                     8764 / 5w3s \cite{hirschi2017cryo} & &\\
 \hline
 9515 / 5gjw \cite{wu2016structure} &
 \multirow{2}{*}{Voltage-gated calcium channel} &
 \multirow{2}{*}{ \ref{fig:maps} d / 3}
                    \\
                     \cline{1-1}
                     6475 / 3jb \cite{wu2015structure} & &\\
 \hline
 
 6284 / 3j9t \cite{zhao2015electron} &
 \multirow{2}{*}{Yeast V-ATPase} &
 \multirow{2}{*}{ \ref{fig:maps} e / 4 }
                    \\
                     \cline{1-1}
                     8724 / 5vox \cite{zhao2017molecular} & &\\
 \hline
 3342 / 5fwm \cite{verba2016atomic} &
 \multirow{2}{*}{Hsp90-Cdc37-Cdk4} &
 \multirow{2}{*}{ \ref{fig:maps} f / 5}
                    \\
                     \cline{1-1}
                     3341 / 5fwl \cite{verba2016atomic} & &\\
 \hline
 
 4547 / 6qg5 \cite{gordiyenko2019structural} &
 \multirow{2}{*}{eIF2B-eIF2} &
 \multirow{2}{*}{\ref{fig:maps} g / 6}
                    \\
                     \cline{1-1}
                     4548 / 6qg6 \cite{gordiyenko2019structural} & &\\
 \hline
\end{tabular}}
\end{center}
\end{table}
\newpage

\begin{table}[h!]
\begin{center}
\caption{Computational performance of \emph{AlignOT}. For point clouds of size $n \in \{50,100,200,500,1000\}$, and for a fixed rotation of $\theta=20^\circ$, we ran \emph{AlignOT} on EMDB:1717 $50$ times. We evaluate the accuracy of the method by computing the angle difference with the ground truth, and the associated runtime in seconds (mean and std). Experiments were run on an Intel(R) Core(TM) workstation with i7-7700HQ CPU @ 2.80GHz   2.81 GHz with 16.0 GB RAM.}
\label{tab:numpoints}
\begin{tabular}{ | c | c | c | }
\hline
$n$ & Angle difference & Runtime (s)\\
\hline
50 & \begin{tabular}[c]{@{}c@{}}$12.60^\circ$$\pm 4.51^\circ$ \end{tabular} & \begin{tabular}[c]{@{}c@{}}$2.36$$\pm 0.42$ \end{tabular}\\

100 & \begin{tabular}[c]{@{}c@{}}$7.72^\circ$$\pm 3.84^\circ$ \end{tabular} & \begin{tabular}[c]{@{}c@{}}$4.66$$\pm 0.57$ \end{tabular}\\

200 & \begin{tabular}[c]{@{}c@{}}$3.85^\circ$$\pm 1.53^\circ$ \end{tabular} & \begin{tabular}[c]{@{}c@{}}$4.14$$\pm 0.37$ \end{tabular}\\

500 & \begin{tabular}[c]{@{}c@{}}$2.35^\circ$$\pm 1.23^\circ$ \end{tabular} & \begin{tabular}[c]{@{}c@{}}$16.36$$\pm 1.37$ \end{tabular}\\

1000 & \begin{tabular}[c]{@{}c@{}}$2.20^\circ$$\pm 0.96^\circ$ \end{tabular} & \begin{tabular}[c]{@{}c@{}}$54.75$$\pm 2.77$ \end{tabular}\\



\hline
\end{tabular}
\end{center}
\end{table}
\newpage

\begin{table}[h!]
\begin{center}
\caption{
Benchmarking of \emph{AlignOT} on pairs of conformationally heterogeneous maps, as listed in Table \ref{tab:dataset2}. For each pair, we ran \emph{AlignOT} and \texttt{fitmap} and reported the angle difference between the resulting algorithm and the ground truth (see Figure \ref{fig:exp2}a). For each pair, we ran our test with initial rotation angles $\theta \in \left\{45,60,90^\circ\right\}$ , across 89 different axes covering half of the sphere $\mathcal{S}^2$ (with 20 runs for each), and point cloud sizes $n \in \left\{250,500,1000\right\}$ in \emph{AlignOT}. Each table entry reports the mean and std observed (with sample size 20), with the best results for each pair highlighted in bold.
To compute the regularized Wasserstein distance of final output for different methods, we sampled $500$ points from each map and used the same regularization parameter $\epsilon=100$.
}
\label{tab:exp_benchmark}
\resizebox{\textwidth}{!}{
\begin{tabular}{ | c | c | c | c | c | c | c | c | c | c | c | }
\hline
    
    
     \multirow{4}{*}{\begin{tabular}[c]{@{}c@{}} ID \end{tabular}} & \multirow{4}{*}{\begin{tabular}[c]{@{}c@{}} Angle \end{tabular}} & \multirow{4}{*}{\begin{tabular}[c]{@{}c@{}}MMAlign \\ $\mathcal{W}_{2,\epsilon}^2$ \end{tabular}}  & \multicolumn{2}{|c|}{\texttt{fitmap}} & \multicolumn{6}{|c|}{\textit{AlignOT}}\\
    
     &  &  & \multicolumn{2}{|c|}{ } & \multicolumn{2}{|c|}{$n=250$} & \multicolumn{2}{|c|}{$n=500$} & \multicolumn{2}{|c|}{$n=1000$}\\
    
     &  &  & \begin{tabular}[c]{@{}c@{}} Angle \\ diff. ($^\circ$) \end{tabular} & $\mathcal{W}_{2,\epsilon}^2$ & \begin{tabular}[c]{@{}c@{}} Angle \\ diff. ($^\circ$) \end{tabular} & $\mathcal{W}_{2,\epsilon}^2$ & \begin{tabular}[c]{@{}c@{}} Angle \\ diff. ($^\circ$) \end{tabular} & $\mathcal{W}_{2,\epsilon}^2$ & \begin{tabular}[c]{@{}c@{}} Angle \\ diff. ($^\circ$) \end{tabular} & $\mathcal{W}_{2,\epsilon}^2$\\
 \hline
 \multirow{5}{*}{\begin{tabular}[c]{@{}c@{}} 1 \end{tabular}} & 45 & \begin{tabular}[c]{@{}c@{}} $125.14$\\$\pm 3.79$  \end{tabular} &  \begin{tabular}[c]{@{}c@{}}$\mathbf{0.97}$\\$\mathbf{\pm0.71}$ \end{tabular} & \begin{tabular}[c]{@{}c@{}}$\mathbf{125.36}$\\$\mathbf{\pm4.15}$\end{tabular}& \begin{tabular}[c]{@{}c@{}}$2.82$\\$\pm1.34$ \end{tabular} & \begin{tabular}[c]{@{}c@{}}$130.73$\\$\pm8.36$ \end{tabular} & \begin{tabular}[c]{@{}c@{}}$2.30$\\$\pm1.08$ \end{tabular} & \begin{tabular}[c]{@{}c@{}}$125.65$\\$\pm5.26$ \end{tabular} & \begin{tabular}[c]{@{}c@{}}$5.73$\\$\pm3.84$ \end{tabular} & \begin{tabular}[c]{@{}c@{}}$127.51$\\$\pm7.27$ \end{tabular}\\
 \cline{2-11}
  & 60 & \begin{tabular}[c]{@{}c@{}} $125.14$\\$\pm 3.79$  \end{tabular} &   \begin{tabular}[c]{@{}c@{}}$6.39$\\$\pm14.35$ \end{tabular} & \begin{tabular}[c]{@{}c@{}}$133.71$\\$\pm 38.01$ \end{tabular} & \begin{tabular}[c]{@{}c@{}}$\mathbf{2.79}$\\$\mathbf{\pm1.32}$ \end{tabular} & \begin{tabular}[c]{@{}c@{}}$130.82$\\$\pm8.42$ \end{tabular} & \begin{tabular}[c]{@{}c@{}}$2.99$\\$\pm1.95$ \end{tabular} & \begin{tabular}[c]{@{}c@{}}$\mathbf{126.19}$\\$\mathbf{\pm5.22}$\end{tabular}& \begin{tabular}[c]{@{}c@{}}$10.78$\\$\pm8.21$ \end{tabular} & \begin{tabular}[c]{@{}c@{}}$140.99$\\$\pm23.44$ \end{tabular} \\
 \cline{2-11}
  & 90 & \begin{tabular}[c]{@{}c@{}} $125.14$\\$\pm 3.79$  \end{tabular} &   \begin{tabular}[c]{@{}c@{}}$46.93$\\$\pm30.70$ \end{tabular} &  \begin{tabular}[c]{@{}c@{}}$267.78$\\$\pm129.91$ \end{tabular} & \begin{tabular}[c]{@{}c@{}}$\mathbf{9.22}$\\$\mathbf{\pm25.99}$\end{tabular} & \begin{tabular}[c]{@{}c@{}}$\mathbf{140.72}$\\$\mathbf{\pm40.58}$\end{tabular}& \begin{tabular}[c]{@{}c@{}}$16.77$\\$\pm22.09$ \end{tabular} & \begin{tabular}[c]{@{}c@{}}$169.43$\\$\pm74.51$ \end{tabular} & \begin{tabular}[c]{@{}c@{}}$32.48$\\$\pm25.48$ \end{tabular} & \begin{tabular}[c]{@{}c@{}}$223.20$\\$\pm94.98$ \end{tabular} \\
 \hline

 \multirow{5}{*}{\begin{tabular}[c]{@{}c@{}} 2 \end{tabular}} & 45 & \begin{tabular}[c]{@{}c@{}} 152.69\\$\pm 9.67$  \end{tabular} & \begin{tabular}[c]{@{}c@{}}$37.91$\\$\pm8.38$ \end{tabular} &  \begin{tabular}[c]{@{}c@{}} $292.44$\\$\pm 30.47$  \end{tabular} & \begin{tabular}[c]{@{}c@{}}$20.01$\\$\pm34.43$ \end{tabular} & \begin{tabular}[c]{@{}c@{}}$188.17$\\$\pm29.67$ \end{tabular} & \begin{tabular}[c]{@{}c@{}}$14.99$\\$\pm30.83$ \end{tabular} & \begin{tabular}[c]{@{}c@{}}$163.25$\\$\pm25.32$ \end{tabular} & \begin{tabular}[c]{@{}c@{}}$\mathbf{11.06}$\\$\mathbf{\pm26.68}$\end{tabular} & \begin{tabular}[c]{@{}c@{}}$\mathbf{149.06}$\\$\mathbf{\pm21.79}$ \end{tabular}\\
 \cline{2-11}
  & 60 & \begin{tabular}[c]{@{}c@{}} 152.69\\$\pm 9.67$  \end{tabular} & \begin{tabular}[c]{@{}c@{}}$\mathbf{55.33}$\\$\mathbf{\pm5.79}$\end{tabular}& \begin{tabular}[c]{@{}c@{}} $317.66$\\$\pm 42.78$  \end{tabular} & \begin{tabular}[c]{@{}c@{}}$62.32$\\$\pm42.37$ \end{tabular} & \begin{tabular}[c]{@{}c@{}}$214.83$\\$\pm42.83$ \end{tabular} & \begin{tabular}[c]{@{}c@{}}$62.88$\\$\pm41.47$ \end{tabular} & \begin{tabular}[c]{@{}c@{}}$195.23$\\$\pm42.07$ \end{tabular} & \begin{tabular}[c]{@{}c@{}}$62.58$\\$\pm39.83$ \end{tabular} & \begin{tabular}[c]{@{}c@{}}$\mathbf{188.42}$\\$\mathbf{\pm45.09}$ \end{tabular} \\
 \cline{2-11}
  & 90 & \begin{tabular}[c]{@{}c@{}} 152.69\\$\pm 9.67$  \end{tabular} &   \begin{tabular}[c]{@{}c@{}}$\mathbf{85.21}$\\$\mathbf{\pm5.42}$\end{tabular} & \begin{tabular}[c]{@{}c@{}} $295.43$\\$\pm 49.50$  \end{tabular} & \begin{tabular}[c]{@{}c@{}}$98.20$\\$\pm21.54$ \end{tabular} & \begin{tabular}[c]{@{}c@{}}$245.20$\\$\pm40.92$ \end{tabular} & \begin{tabular}[c]{@{}c@{}}$99.69$\\$\pm14.84$ \end{tabular} & \begin{tabular}[c]{@{}c@{}}$228.26$\\$\pm39.47$ \end{tabular} & \begin{tabular}[c]{@{}c@{}}$99.66$\\$\pm13.48$ \end{tabular} & \begin{tabular}[c]{@{}c@{}}$\mathbf{221.37}$\\$\mathbf{\pm40.41}$\end{tabular} \\
 \hline

 \multirow{5}{*}{\begin{tabular}[c]{@{}c@{}} 3 \end{tabular}} & 45 & \begin{tabular}[c]{@{}c@{}} 237.78\\$\pm 31.97$ \end{tabular}  & \begin{tabular}[c]{@{}c@{}}$\mathbf{3.56}$\\$\mathbf{\pm10.48}$\end{tabular} & \begin{tabular}[c]{@{}c@{}} $268.36$\\$\pm 147.22$ \end{tabular} & \begin{tabular}[c]{@{}c@{}}$6.71$\\$\pm10.13$ \end{tabular} & \begin{tabular}[c]{@{}c@{}}$263.32$\\$\pm46.31$ \end{tabular} & \begin{tabular}[c]{@{}c@{}}$5.41$\\$\pm1.66$ \end{tabular} & \begin{tabular}[c]{@{}c@{}}$221.49$\\$\pm22.01$ \end{tabular} & \begin{tabular}[c]{@{}c@{}}$5.14$\\$\pm1.20$ \end{tabular} & \begin{tabular}[c]{@{}c@{}}$\mathbf{201.35}$\\$\mathbf{\pm11.38}$\end{tabular}\\
 \cline{2-11}
  & 60 & \begin{tabular}[c]{@{}c@{}} 237.78\\$\pm 31.97$ \end{tabular} & \begin{tabular}[c]{@{}c@{}} $38.35$\\$\pm 20.65$  \end{tabular} & \begin{tabular}[c]{@{}c@{}} $823.87$\\$\pm 415.12$ \end{tabular} & \begin{tabular}[c]{@{}c@{}}$9.49$\\$\pm24.21$ \end{tabular} & \begin{tabular}[c]{@{}c@{}}$267.61$\\$\pm64.31$ \end{tabular} & \begin{tabular}[c]{@{}c@{}}$7.00$\\$\pm16.86$ \end{tabular} & \begin{tabular}[c]{@{}c@{}}$224.56$\\$\pm44.38$ \end{tabular} & \begin{tabular}[c]{@{}c@{}}$\mathbf{5.13}$\\$\mathbf{\pm1.20}$ \end{tabular} & \begin{tabular}[c]{@{}c@{}}$\mathbf{201.86}$\\$\mathbf{\pm12.05}$\end{tabular} \\
 \cline{2-11}
  & 90 & \begin{tabular}[c]{@{}c@{}} 237.78\\$\pm 31.97$ \end{tabular} & \begin{tabular}[c]{@{}c@{}} $80.53$\\$\pm 7.75$  \end{tabular} & \begin{tabular}[c]{@{}c@{}} $1623.24$\\$\pm 521.41$ \end{tabular} & \begin{tabular}[c]{@{}c@{}}$37.25$\\$\pm66.03$ \end{tabular} &  \begin{tabular}[c]{@{}c@{}}$330.59$\\$\pm157.46$ \end{tabular} & \begin{tabular}[c]{@{}c@{}}$30.94$\\$\pm61.25$ \end{tabular} & \begin{tabular}[c]{@{}c@{}}$281.16$\\$\pm144.28$ \end{tabular} & \begin{tabular}[c]{@{}c@{}}$\mathbf{25.27}$\\$\mathbf{\pm55.53}$\end{tabular}& \begin{tabular}[c]{@{}c@{}}$\mathbf{248.83}$\\$\mathbf{\pm130.86}$\end{tabular} \\
 \hline

 \multirow{5}{*}{\begin{tabular}[c]{@{}c@{}} 4 \end{tabular}} & 45 & \begin{tabular}[c]{@{}c@{}}413.84\\$\pm 35.57$ \end{tabular}  & \begin{tabular}[c]{@{}c@{}}$31.65$\\$\pm 10.33$  \end{tabular} & \begin{tabular}[c]{@{}c@{}}$\mathbf{325.64}$\\$\mathbf{\pm 83.43}$\end{tabular} & \begin{tabular}[c]{@{}c@{}}$10.45$\\$\pm2.67$ \end{tabular} &  \begin{tabular}[c]{@{}c@{}}$422.41$\\$\pm44.29$ \end{tabular} & \begin{tabular}[c]{@{}c@{}}$10.05$\\$\pm1.94$ \end{tabular} & \begin{tabular}[c]{@{}c@{}}$408.48$\\$\pm30.38$ \end{tabular} & \begin{tabular}[c]{@{}c@{}}$\mathbf{9.65}$\\$\mathbf{\pm1.49}$ \end{tabular} & \begin{tabular}[c]{@{}c@{}}$401.14$\\$\pm21.32$ \end{tabular}\\
 \cline{2-11}
  & 60 & \begin{tabular}[c]{@{}c@{}}413.84\\$\pm 35.57$ \end{tabular} & \begin{tabular}[c]{@{}c@{}} $47.21$\\$\pm 11.27$  \end{tabular} & \begin{tabular}[c]{@{}c@{}}$435.61$\\$\pm 126.95$ \end{tabular} & \begin{tabular}[c]{@{}c@{}}$10.75$\\$\pm6.04$ \end{tabular} &  \begin{tabular}[c]{@{}c@{}}$422.46$\\$\pm42.70$ \end{tabular} &  \begin{tabular}[c]{@{}c@{}}$10.07$\\$\pm4.09$ \end{tabular} & \begin{tabular}[c]{@{}c@{}}$407.52$\\$\pm30.02$ \end{tabular} & \begin{tabular}[c]{@{}c@{}}$\mathbf{9.56}$\\$\mathbf{\pm1.51}$ \end{tabular} & \begin{tabular}[c]{@{}c@{}}$\mathbf{401.28}$\\$\mathbf{\pm21.67}$ \end{tabular} \\
 \cline{2-11}
  & 90 & \begin{tabular}[c]{@{}c@{}}413.84\\$\pm 35.57$ \end{tabular} & \begin{tabular}[c]{@{}c@{}} $81.79$\\$\pm 6.69$  \end{tabular} & \begin{tabular}[c]{@{}c@{}}$655.52$\\$\pm 170.82$ \end{tabular} & \begin{tabular}[c]{@{}c@{}}$33.51$\\$\pm57.08$ \end{tabular} &  \begin{tabular}[c]{@{}c@{}}$440.99$\\$\pm65.55$ \end{tabular} &  \begin{tabular}[c]{@{}c@{}}$29.54$\\$\pm52.75$ \end{tabular} & \begin{tabular}[c]{@{}c@{}}$\mathbf{425.56}$\\$\mathbf{\pm60.30}$\end{tabular} & \begin{tabular}[c]{@{}c@{}}$\mathbf{26.65}$\\$\mathbf{\pm46.27}$\end{tabular} & \begin{tabular}[c]{@{}c@{}}$432.05$\\$\pm85.33$ \end{tabular}  \\
 \hline

 \multirow{5}{*}{\begin{tabular}[c]{@{}c@{}} 5 \end{tabular}} & 45 & \begin{tabular}[c]{@{}c@{}} 104.93\\$\pm 1.53$  \end{tabular} & \begin{tabular}[c]{@{}c@{}} $2.02$\\$\pm 5.41$  \end{tabular} & \begin{tabular}[c]{@{}c@{}}$111.62$\\$\pm 15.84$ \end{tabular} & \begin{tabular}[c]{@{}c@{}}$5.94$\\$\pm15.79$ \end{tabular} & \begin{tabular}[c]{@{}c@{}}$106.87$\\$\pm3.95$ \end{tabular} & \begin{tabular}[c]{@{}c@{}}$3.77$\\$\pm11.68$ \end{tabular} & \begin{tabular}[c]{@{}c@{}}$104.93$\\$\pm2.66$ \end{tabular} & \begin{tabular}[c]{@{}c@{}}$\mathbf{2.02}$\\$\mathbf{\pm1.22}$ \end{tabular} & \begin{tabular}[c]{@{}c@{}}$\mathbf{103.98}$\\$\mathbf{\pm1.32}$\end{tabular}\\
 \cline{2-11}
  & 60 & \begin{tabular}[c]{@{}c@{}} 104.93\\$\pm 1.53$  \end{tabular} &   \begin{tabular}[c]{@{}c@{}} $26.10$\\$\pm 19.97$  \end{tabular} & \begin{tabular}[c]{@{}c@{}}$138.80$\\$\pm 30.50$ \end{tabular} & \begin{tabular}[c]{@{}c@{}}$9.37$\\$\pm28.47$ \end{tabular} & \begin{tabular}[c]{@{}c@{}}$107.42$\\$\pm5.51$ \end{tabular} & \begin{tabular}[c]{@{}c@{}}$6.64$\\$\pm24.80$ \end{tabular} & \begin{tabular}[c]{@{}c@{}}$105.38$\\$\pm4.53$ \end{tabular} & \begin{tabular}[c]{@{}c@{}}$\mathbf{3.37}$\\$\mathbf{\pm14.33}$\end{tabular} & \begin{tabular}[c]{@{}c@{}}$\mathbf{104.35}$\\$\mathbf{\pm3.80}$\end{tabular}  \\
 \cline{2-11}
  & 90 & \begin{tabular}[c]{@{}c@{}} 104.93\\$\pm 1.53$  \end{tabular} &   \begin{tabular}[c]{@{}c@{}} $60.52$\\$\pm 16.78$  \end{tabular} & \begin{tabular}[c]{@{}c@{}}$162.30$\\$\pm 23.35$ \end{tabular} & \begin{tabular}[c]{@{}c@{}}$17.454$\\$\pm45.22$ \end{tabular} & \begin{tabular}[c]{@{}c@{}}$108.72$\\$\pm7.94$ \end{tabular} & \begin{tabular}[c]{@{}c@{}}$26.27$\\$\pm58.88$ \end{tabular} & \begin{tabular}[c]{@{}c@{}}$108.56$\\$\pm9.91$ \end{tabular} & \begin{tabular}[c]{@{}c@{}}$\mathbf{16.71}$\\$\mathbf{\pm47.56}$\end{tabular} & \begin{tabular}[c]{@{}c@{}}$\mathbf{106.79}$\\$\mathbf{\pm9.03}$\end{tabular} \\
 \hline

 \multirow{5}{*}{\begin{tabular}[c]{@{}c@{}} 6 \end{tabular}} & 45 & \begin{tabular}[c]{@{}c@{}} 119.92\\$\pm 2.91$  \end{tabular} & \begin{tabular}[c]{@{}c@{}}$\mathbf{0.91}$\\$\mathbf{\pm4.98}$ \end{tabular} & \begin{tabular}[c]{@{}c@{}}$\mathbf{113.25}$\\$\mathbf{\pm16.73}$\end{tabular}& \begin{tabular}[c]{@{}c@{}}$4.26$\\$\pm1.58$ \end{tabular} & \begin{tabular}[c]{@{}c@{}}$120.77$\\$\pm5.34$ \end{tabular} & \begin{tabular}[c]{@{}c@{}}$7.54$\\$\pm4.20$ \end{tabular} & \begin{tabular}[c]{@{}c@{}}$122.15$\\$\pm5.49$ \end{tabular} & \begin{tabular}[c]{@{}c@{}}$14.53$\\$\pm6.93$ \end{tabular} & \begin{tabular}[c]{@{}c@{}}$138.26$\\$\pm11.07$ \end{tabular}\\
 \cline{2-11}
  & 60 & \begin{tabular}[c]{@{}c@{}} 119.92\\$\pm 2.91$  \end{tabular} & \begin{tabular}[c]{@{}c@{}}$14.73$\\$\pm19.64$ \end{tabular} & \begin{tabular}[c]{@{}c@{}}$155.34$\\$\pm 58.90$ \end{tabular} & \begin{tabular}[c]{@{}c@{}}$\mathbf{5.17}$\\$\mathbf{\pm2.33}$ \end{tabular} & \begin{tabular}[c]{@{}c@{}}$\mathbf{121.43}$\\$\mathbf{\pm5.52}$\end{tabular}& \begin{tabular}[c]{@{}c@{}}$11.50$\\$\pm7.26$ \end{tabular} & \begin{tabular}[c]{@{}c@{}}$128.46$\\$\pm11.18$ \end{tabular} & \begin{tabular}[c]{@{}c@{}}$22.12$\\$\pm10.46$ \end{tabular} & \begin{tabular}[c]{@{}c@{}}$161.75$\\$\pm20.83$ \end{tabular} \\
 \cline{2-11}
  & 90 & \begin{tabular}[c]{@{}c@{}} 119.92\\$\pm 2.91$  \end{tabular} &  \begin{tabular}[c]{@{}c@{}}$43.43$\\$\pm27.87$ \end{tabular} & \begin{tabular}[c]{@{}c@{}}$204.40$\\$\pm 57.11$ \end{tabular} & \begin{tabular}[c]{@{}c@{}}$\mathbf{18.37}$\\$\mathbf{\pm26.30}$\end{tabular}& \begin{tabular}[c]{@{}c@{}}$\mathbf{143.74}$\\$\mathbf{\pm45.67}$\end{tabular} & \begin{tabular}[c]{@{}c@{}}$31.68$\\$\pm25.94$ \end{tabular} & \begin{tabular}[c]{@{}c@{}}$178.16$\\$\pm60.72$ \end{tabular} & \begin{tabular}[c]{@{}c@{}}$46.43$\\$\pm22.89$ \end{tabular} & \begin{tabular}[c]{@{}c@{}}$258.30$\\$\pm45.43$ \end{tabular} \\
 \hline

\end{tabular}}
\end{center}
\end{table}
\newpage

\section*{Appendix}
\appendix
    \section*{Appendix A: Optimal translation for the rigid body alignment problem}\label{sec:translation}
        To minimize the loss function 
        over rigid body transformations (translation and rotations), we can simplify the problem and restrict the search to rotations only as follows: For two point clouds $\mathbf{A} = \{a_1,\dots,a_n\}$ and $\mathbf{B} = \{b_1, \dots, b_n\}$, and 
        upon introducing the centers of mass $\Bar{a} = \frac{1}{n}\sum_{i=1}^n a_i$ and $\Bar{b} = \frac{1}{n}\sum_{i=1}^n b_i$, and the centered coordinates $a_{c_i} = a_i - \Bar{a}$ and $b_{c_i} = b_i - \Bar{b}$, equation \eqref{eq:loss_def} 
        yields 
        
        \begin{align}
        \mathcal{L}(R, T) &= \mathcal{W}_{2,\epsilon}(\text{move}_{R,T}(\mathbf{A}),\mathbf{B})^2 \nonumber\\
        &= \min_{P \in \mathbb{R}_+^{n \times n}} \sum_{i,j=1}^n \|Ra_i+T-b_j\|_2^2P_{i, j} + \epsilon H(P)   \quad(\textrm{s.t.} \quad  \forall 1 \le j \le n: \sum_{i=1}^n P_{i,j} = \sum_{i=1}^n P_{j,i} = \frac{1}{n}) \nonumber\\
        &= \min_{P \in \mathbb{R}_+^{n \times n}} \sum_{i,j=1}^n \|Ra_{c_i} + R\Bar{a} + T - b_{c_j} - \Bar{b}\|_2^2P_{i, j} + \epsilon H(P), \nonumber
        \end{align}
          where we used definition of the entropy regularized Wasserstein distance from equation  \eqref{eq:Wass_dist}. 
          This further simplifies as
         \begin{align}
        \mathcal{L}(R, T) &= \min_{P \in \mathbb{R}_+^{n \times n}} \sum_{i,j=1}^n \|Ra_{c_i} - b_{c_j}\|_2^2P_{i, j}  + \sum_{i,j=1}^n\|R\Bar{a} + T - \Bar{b}\|_2^2P_{i, j} \nonumber\\& \quad \; + \sum_{i,j=1}^n(Ra_{c_i} - b_{c_j}).(R\Bar{a} + T - \Bar{b})P_{i, j} + \epsilon H(P)\label{eq:loss2}\\
        &= \min_{P \in \mathbb{R}_+^{n \times n}} \sum_{i,j=1}^n \|Ra_{c_i} - b_{c_j}\|_2^2P_{i, j}   + \sum_{i,j=1}^n\|R\Bar{a} + T - \Bar{b}\|_2^2P_{i, j} + \epsilon H(P)\label{eq:clear}\\
        &= \min_{P \in \mathbb{R}_+^{n \times n}} [\sum_{i,j=1}^n \|Ra_{c_i} - b_{c_j}\|_2^2P_{i, j} + \epsilon H(P)]  + \|R\Bar{a} + T - \Bar{b}\|_2^2\label{eq:loss3}
        \end{align}
        

        where we used in \eqref{eq:loss2} that $\sum_{i,j=1}^n P_{i,j}a_{c_i} = \frac{1}{n}\sum_{i=1}^n a_{c_i} =0$ and $\sum_{i,j=1}^n P_{i,j}b_{c_j} = \frac{1}{n}\sum_{j=1}^n b_{c_j} = 0$. 
        Also, we used the fact that $\sum_{i,j=1}^n P_{i,j} = 1$ in \eqref{eq:clear}.
        The second term in eq. \eqref{eq:loss3} is minimized for $T = \Bar{b} - R \Bar{a}$, i.e. the translation that aligns the two centers of mass. Therefore, to solve the rigid body alignment problem, we can assume  that the distributions of points $\mathbf{A}$ and $\mathbf{B}$ are both centered at the origin, and that we only need to solve the optimization problem over the rotations in $SO(3)$, as stated in equation \eqref{eq:minOT_new}.

\section*{Appendix B: Quaternion Representation of 3D rotations} \label{sec:quaternions}

\subsection*{Quaternion Representation of 3D rotations and gradient formula}

    To formalize 3D rotations in \textit{AlignOT} we use the quaternion representation ($\mathbb{H}$).
   In this representation,
   given a point $a = (x,y,z) \in \mathbb{R}^3$ 
   and a rotation with angle $\theta$ around axis $\vec{u} = (u_x,u_y,u_z)$
   , we form quaternions
        \begin{align}
        q &= \cos \theta/2 + u_x \sin \theta/2 i+ u_y \sin \theta/2 j+ u_z \sin \theta/2k,\nonumber\\
        p &= xi + yj + zk, \nonumber
        \end{align}
         where $i,j,k$ are the basic quaternions such that
        \begin{equation}
            i^2=j^2=k^2=ijk=-1. \label{eq:mul_gol}
        \end{equation}
        Using equation \eqref{eq:mul_gol} we compute the following term
        \begin{equation}
            x_{\text{res}}i + y_{\text{res}}j + z_{\text{res}}k = q  p  q^*,\nonumber
        \end{equation}
        where $R_q(a) = (x_\text{res},y_\text{res},z_\text{res}) \in \mathbb{R}^3$ is the coordinates of $a$ after rotation, and $q^*$ is the conjugate of $q$ and is defined as
        \begin{equation}
            (q_0 + q_1 i + q_2 j + q_3k)^* = q_0 - q_1 i - q_2 j - q_3k.\nonumber
        \end{equation}
         We then define the absolute norm of $q$ as
        \begin{equation}
            \|q\|^2 = q q^* = q_0^2 + q_1^2 + q_2^2 + q_3^2 \in \mathbb{R}.\nonumber
        \end{equation}
        
        In \textit{AlignOT}, we also compute the gradient associated with the function $f(q) = f(q_0,q_1,q_2,q_3) = \| R_q(a) - b \|_2^2 $, where $a,b\in \mathbb{R}^3$ as
     
        $$\nabla f = \frac{\partial f}{\partial q_0}  + \frac{\partial f}{\partial q_1}i  + \frac{\partial f}{\partial q_2}j  + \frac{\partial f}{\partial q_3}k,$$
        where the explicit analytical formula  of $\nabla f$ is provided in the Appendix B.

\subsection*{Analytical formula for the gradient update}
In the previous section we defined gradient of a general function $f: \mathbb{H}\rightarrow \mathbb{R}$. In this section
use this definition to compute an explicit formula for the gradient of the function used in Algorithm 1 i.e. $f(q) = d(R_q(a),b)^2$, where $d$ is Euclidean distance in $\mathbb{R}^3$, $a=(a_x,a_y,a_z),b=(b_x,b_y,b_z)$ are two points in $\mathbb{R}^3$, and $q = q_0 + q_1i + q_2j + q_3k$ is a quaternion. Using the definition of $R_q(a)$ in Appendix B to compute the coordinates of $R_q(a) = (a_{x_{res}}, a_{y_{res}}, a_{z_{res}})$, we have
$$a_{x_{res}}i + a_{y_{res}}j + a_{z_{res}}k = q(a_xi+a_yj+a_zk)q^*=f_i(q)i + f_j(q)j + f_k(q)k.$$
After some simplification, we can write $f_i,f_j$ and $f_k$ as
\begin{align}
a_{x_{res}} = f_i(q)=q_1^2a_x + 2q_1q_2a_y + 2q_1q_3a_z + q_0^2a_x + 2q_0q_2a_z - 2q_0q_3a_y - q_3^2a_x - q_2^2a_x,\nonumber\\
a_{y_{res}} = f_j(q) = 2q_1q_2a_x + q_2^2a_y + 2q_2q_3a_z + q_0^2a_y + 2q_0q_3a_x - 2q_0q_1a_z - q_1^2a_y - q_3^2a_y,\nonumber\\
a_{z_{res}} = f_k(q) = 2q_1q_3a_x + 2q_2q_3a_y + q_3^2a_z + q_0^2a_z + 2q_0q_1a_y - 2q_0q_2a_x - q_2^2a_z - q_1^2a_z.\nonumber
\end{align}
To compute partial derivatives of $f$ we use that
\begin{align}
\frac{\partial f}{\partial q_l} &= \frac{(a_{x_{res}} - b_x)^2 + (a_{y_{res}} - b_y)^2 + (a_{z_{res}} - b_z)^2}{\partial q_l}\nonumber\\
&= 2(f_i(q) - b_x)\frac{\partial f_i(q)}{\partial q_l} +  2(f_j(q) - b_y)\frac{\partial f_j(q)}{\partial q_l} +  2(f_k(q) - b_z)\frac{\partial f_k(q)}{\partial q_l}\nonumber
\end{align}
for $l \in \{0,1,2,3\}$, which yields
\begin{align}
\frac{\partial f}{\partial q_0} &=  2(f_i(q) - b_x)\frac{\partial f_i(q)}{\partial q_0} +  2(f_j(q) - b_y)\frac{\partial f_j(q)}{\partial q_0} +  2(f_k(q) - b_z)\frac{\partial f_k(q)}{\partial q_0}\nonumber\\
&=  4(f_i(q) - b_x)(q_0a_x + q_2a_z - q_3a_y) +  4(f_j(q) - b_y)(q_0a_y + q_3a_x - q_1a_z) \nonumber\\
& \quad  +4(f_k(q) - b_z)(q_0a_z + q_1a_y - q_2a_x),\nonumber\\
\frac{\partial f}{\partial q_1} &=  2(f_i(q) - b_x)\frac{\partial f_i(q)}{\partial q_1} +  2(f_j(q) - b_y)\frac{\partial f_j(q)}{\partial q_1} +  2(f_k(q) - b_z)\frac{\partial f_k(q)}{\partial q_1}\nonumber\\
&=  4(f_i(q) - b_x)(q_1a_x + q_2a_y + q_3a_z) +  4(f_j(q) - b_y)(q_2a_x - q_0a_z - q_1a_y) \nonumber\\
& \quad +4(f_k(q) - b_z)(q_3a_x + q_0a_y - q_1a_z),\nonumber\\
\frac{\partial f}{\partial q_2} &=  2(f_i(q) - b_x)\frac{\partial f_i(q)}{\partial q_2} +  2(f_j(q) - b_y)\frac{\partial f_j(q)}{\partial q_2} +  2(f_k(q) - b_z)\frac{\partial f_k(q)}{\partial q_2}\nonumber\\
&=  4(f_i(q) - b_x)(q_1a_y + q_0a_z - q_2a_x) +  4(f_j(q) - b_y)(q_1a_x + q_2a_y + q_3a_z) \nonumber\\
& \quad +4(f_k(q) - b_z)(q_3a_y - q_0a_x - q_2a_z),\nonumber\\
\frac{\partial f}{\partial q_3} &=  2(f_i(q) - b_x)\frac{\partial f_i(q)}{\partial q_3} +  2(f_j(q) - b_y)\frac{\partial f_j(q)}{\partial q_3} +  2(f_k(q) - b_z)\frac{\partial f_k(q)}{\partial q_3}\nonumber\\
&=  4(f_i(q) - b_x)(q_1a_z - q_0a_y - q_3a_x) +  4(f_j(q) - b_y)(q_2a_z + q_0a_x - q_3a_y)\nonumber \\
& \quad +4(f_k(q) - b_z)(q_1a_x + q_2a_y + q_3a_z),\nonumber\\
\nabla f &=  \frac{\partial f}{q_0} + \frac{\partial f}{q_1}i + \frac{\partial f}{q_2}j + \frac{\partial f}{q_3}k.\nonumber
\end{align}

\clearpage

\end{document}